\documentclass[preprint]{aastex}

\newcommand{\eg}{\emph{e.g.}}
\newcommand{\ie}{\emph{i.e.}}

\newcommand{\myemail}{berian@berkeley.edu}

\shorttitle{Non-Gaussian fields and the genus statistic}
\shortauthors{James}

\begin{document}


\title{Physics of non-Gaussian fields and the cosmological genus statistic}


\author{J. Berian James}
\affil{Astronomy Department, University of California, Berkeley, CA 94720 \and Dark Cosmology Centre, University of Copenhagen,\\ Juliane Maries Vej 30, 2100 Copenhagen \O, Denmark.}
\email{\myemail}


\begin{abstract}
We report a technique to calculate the impact of distinct physical processes inducing non-Gaussianity on the cosmological density field. A natural decomposition of the cosmic genus statistic into an orthogonal polynomial sequence allows complete expression of the scale-dependent evolution of the topology of large-scale structure, in which effects including galaxy bias, nonlinear gravitational evolution and primordial non-Gaussianity may be delineated. The relationship of this decomposition to previous methods for analysing the genus statistic is briefly considered and the following applications are made: i) the expression of certain systematics affecting topological measurements; ii) the quantification of broad deformations from Gaussianity that appear in the genus statistic as measured in the Horizon Run simulation; iii) the study of the evolution of the genus curve for simulations with primordial non-Gaussianity. These advances improve the treatment of flux-limited galaxy catalogues for use with this measurement and further the use of the genus statistic as a tool for exploring non-Gaussianity.
\end{abstract}


\keywords{cosmology: theory---dark matter---methods: statistical}

\section{Introduction}

The study of the topology of large-scale structure promises much from the current generation of galaxy redshift surveys. On the largest scales, the relationship between the development of structure and primordial non-Gaussianity has recently deepened; in the regime of nonlinear structure formation, the effects of gravitational evolution and the complicated phenomenon of bias continue to stimulate interest in the fields of cosmology and galactic astrophysics, where, however, the use of the (geometric) correlation function still predominates. The reason for this seems to be that the topology of large-scale structure is difficult to interpret when the underlying distribution departs from a Gaussian random field~\citep{2005ApJ...633...11P,2007MNRAS.375..128J,2006astro.ph.10762G,2009MNRAS.394..454J,Choi10}.

Of the four Minkowski functionals that characterize the morphology of the cosmological density field, the genus statistic of \citet{GMD} has been most frequently employed. Given a choice of critical density value, the genus is a measure of connection or isolation of structures enclosed by the surface of that constant critical density. Different choices of density threshold excise filaments, cluster and voids, whose topological physics is then expressed in compact form. The curve of genus as a function of density has a known analytical form when the underlying density field is of Gaussian random form---the task is to understand the manifestation of non-Gaussian physics in the topology of cosmic structure.

\subsection{Regarding density thresholds}
To help with the development of these ideas, several prevailing definitions of density thresholds are reconciled. The parameterisation of density used in genus measurements is idiosyncratic, but well-motivated; it is the simplest mapping independent of the underlying density distribution function. There are two (less preferable) alternatives: for density fields $\delta = \rho/\bar\rho - 1$ whose distribution functions $f(\delta)$ depart from Gaussian form---as the cosmological density field must over at least some scales---the parameter $\nu = \delta/\sigma_\rho$ is a misleading mapping, imposing as it does an artificial symmetry about mean density. The often-mentioned lognormal model for the distribution function of \citet{1991MNRAS.248....1C} improves this insofar as it corresponds to the distribution function over a broader range of scales: using the parameter $\tilde\delta = \ln(1+\delta)$, the mapping is given by $\nu = \tilde\delta / \sigma_{\tilde\delta}$. Except in highly evolved fields, this mapping is nearly equivalent to the most general parameterisation:
\begin{equation} \nu \equiv \frac{f_G(\delta)-\bar{f_G}}{\sigma_{f_G}}\textrm{, where } f_G(\delta)\equiv\mathrm{Erf}^{-1}\left[\int_{-\infty}^\delta f(\delta')d\delta'\right]; \end{equation}
and it is this variable that represents density in topological statistics. This choice of mapping insists that regions should be judged as comparably over- and under-dense when they occupy the same volume, rather than on the basis of their density.

\section{A Hermite function transform}
The genus of a two-dimensional surface is a measure of its connectedness: the number of holes through the surface less the number of isolated regions (plus 1, by convention); a donut has genus 1, a sphere 0 and two spheres $-1$. The genus curve of a Gaussian random field, and that of the primordial Universe in the simplest models of inflation, is understood to have the form~\citep{BST}
\begin{equation}
 g_\mathrm{GRF}(\nu) \propto \exp\left(-\frac{\nu^2}{2}\right)\left(1-\nu^2\right),\label{eq:grf}
\end{equation}
with the constant of proportionality a functional of the power spectrum of the field; in any event, it is the shape of the curve that is to be scrutinized. 

The curve of genus number as a function of density provides information even in this extremely restricted case: structures separating the highest-density half of the Universe from the lowest-density half are many times connected; excising extreme densities yields many disjoint regions. The balance between these two regimes and the relative abundance of isolated high and low density regions are a scale-dependent probe of many physical processes. Consequently it is desirable to study the surface of genus number as a function of density and scale. In practice, different scales are analysed separately before being compared.

To address the manner in which distinct physical processes alter the topology of large-scale structure, it is proposed that the genus surface (\ie, the genus curve as a function of scale, characterized by Gaussian smoothing length $\lambda$) be decomposed in an orthogonal basis of Hermite functions:
\begin{eqnarray}
 g(\nu;\lambda) & = & \sum_{n=0}^\infty a_n(\lambda)\psi_n(\nu) \\
 \Leftrightarrow a_n(\lambda) & = & \int_{-\infty}^\infty g(\nu;\lambda)\psi_n(\nu)d\nu. \label{eq:decomposition}
\end{eqnarray}
The Hermite functions $\psi_n(\nu)$ are themselves weighted analogues of the Hermite polynomials\footnote{There are two definitions for the Hermite polynomials, conventionally referred to as the `physicist' and `probabilist' forms. The probabilists' definition is used in this communication, as the genus curve for a Gaussian random field is more naturally expressed this way.}
\begin{eqnarray}
 H_n(\nu) & = & (-1)^n e^{\nu^2/2}\left(\frac{d}{d\nu}\right)^n e^{-\nu^2/2} \\
 \psi_n(\nu) & = & \frac{1}{\sqrt{n!\sqrt{2\pi}}} e^{-\nu^2/4} H_n(\nu)
 \end{eqnarray}
which will satisfy the orthogonality relation
\begin{equation}
 \int_{-\infty}^\infty \psi_m(\nu)\psi_n(\nu) d\nu = \delta_{mn}
\end{equation}
required for the decomposition~(\ref{eq:decomposition}) to exist. The evolution of the Hermite modes encode, as a function of scale, the imprint of the physical processes that have modified the field from the Gaussian random form. The sequence of coefficients $a_n$ is referred to as the Hermite spectrum of the genus curve. The normalised sequence of coefficients, $\tilde a_n = |a_n| / \sum_m |a_m|$ is also employed throughout this work as it provides a satisfying interpretation of the fractional contribution of each Hermite mode as a probability distribution across the modes, with different physical processes mapping onto different distributions. In particular, the quantity $\tilde a_2$ is a completely general measure of the Gaussianity in the field: for a Gaussian random field, $\tilde a_2=1$; for all non-Gaussian fields, $\tilde a_2 < 1$, where the particulars of the non-Gaussian process are encoded in the other coefficients.

Much of the arithmetic is simplified by the rescaling $g(\nu)\rightarrow e^{\nu^2/4}g(\nu)$, which ensures that any Gaussian random field has a trivial decomposition: in this scheme, the base mode of the transform is $n=2$, \ie, $g_\mathrm{GRF}(\nu) \propto \psi_2(\nu)$, with other low modes appearing as the field is perturbed from a Gaussian state. Odd-numbered modes will, in particular, introduce asymmetric features corresponding to an overabundance of either isolated clusters or voids. In practice, this means that a measured genus curve should be multiplied by a factor of $e^{\nu^2/4}$ before the decomposition, a convention  adopted throughout this work.

Quantifying the distortions to the genus curve due to a physical process inducing non-Gaussianity has long been understood to deserve attention. Two approaches that have led to progress are: i) the analytical method invoking second-order perturbation theory to calculate the impact of small departures from Gaussianity; and ii) the use of a phenomenological set of genus-related statistics that quantify differences between a measured genus curve and that for a Gaussian random field. In the following sections, the formalism described above is applied to the examination of a range of physical and unphysical (\emph{i.e.}, systematic effects) phenomena that create non-Gaussian signatures in the cosmological density field.


\section{Application I: Formulae for treatment of density field systematics}

In this section, the formalism is applied to the task of correcting observational measurements of the genus statistic for systematic selection effects.  These different effects distort the genus curve in a non-trivial way, to which correction formulae are derived as a function of the relevant reconstruction parameter. These correction formulae make use of the Hermite decomposition technique for genus analysis.

To examine these systematics, use is made of simulated Gaussian random fields of power spectrum $P(k)\propto k^n$, though this choice of function is arbitrary up to the amplitude of the resulting genus curve. The fields are constructed in a spherical polar Fourier space at each wavenumber $k$ by assigning Fourier amplitudes from a Rayleigh distribution of parameter $P(k)$, phases from a distribution uniform in the interval $[0,2\pi)$ and applying the conjugate (Hermitian) symmetry required to ensure a real-value configuration-space density field.

\subsection{Sampling systematic}
To return a fair reconstruction of the density field from a density-dependent sampling (\emph{i.e.}, $\bar{n}\propto\rho\Rightarrow b\sim1$) of $N$ points, it is necessary to smooth the field with a filter exceeding a characteristic length scale of the point distribution. When this condition is not satisfied, the density field is not adequately reconstructed at that scale, and the genus measurement of the smoothed field will not be accurate. There is excellent motivation to understand and correct this effect, as doing so improves the utility of flux-limited samples for the purpose of the genus measurement, offering order-of-magnitude improvements in signal relative to a volume-limited-sample. This subsection provides a parametric form for this correction and demonstrates its application to the case of a Gaussian random field with power-law power spectrum.

To examine this systematic, the same Gaussian random field realisation is Poisson sampled $n_r=100$ times. The variance associated with the correction to the statistic is not explored in this work. Each sample is smoothed through a Gaussian window of scale parameter $\lambda$ and the genus curve for the recovered field is measured. The quantity used here to parameterise the distortion present in the genus curve is the ratio of a characteristic length scale in the point distribution to the smoothing length:
\begin{equation} \Upsilon \equiv \frac{\lambda}{\bar\ell};\end{equation}
the quantity $\bar\ell$ can admit a variety of definitions, the choice of which is only important in so far as it be treated consistently. Among these are the radius of the average spherical volume per point; the side length of the average Euclidean volume per point; the empirical average separation between every pair of points; and the average nearest-neighbour separation of a Poisson distribution with the same sampling density;
indeed, it seems that the definitions return length scales that are equivalent up to a constant factor. For the purpose of this work, the first of these
\begin{equation} \bar\ell \equiv \left(\frac{3V}{4\pi N}\right)^{1/3}\end{equation}
is chosen.

\begin{figure}
\centering
\includegraphics[width=\linewidth]{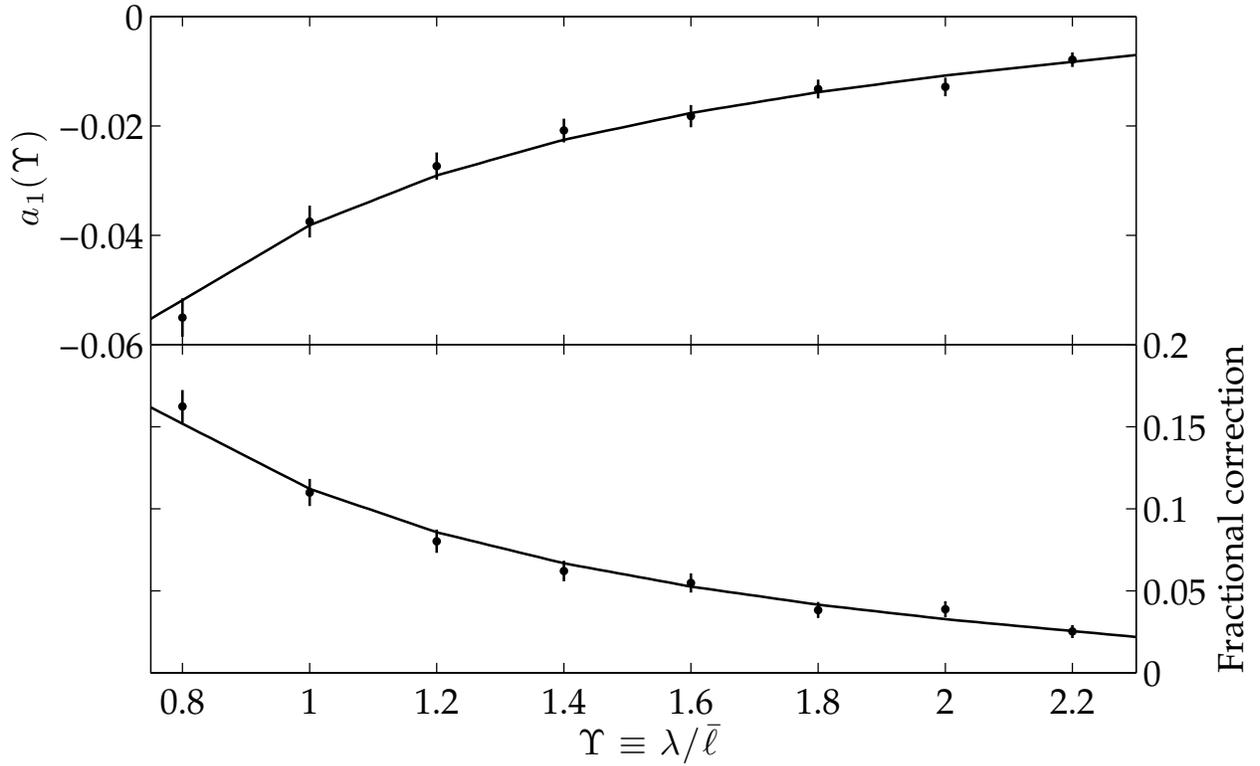}
\caption{Sampling systematic correction factor for a point-sampled Gaussian random field: as the smoothing length falls below the mean separation between points, the asymmetry in the genus statistic is accurately corrected by the first-order Hermite function with amplitude $a_1(\Upsilon)$ (top panel); the magnitude of this correction relative to the amplitude of the base function, the second-order Hermite function, is shown as the fractional correction (bottom panel). This example shown here is for a field of power-law power spectrum with index $n=2$.}\label{fig:sampling}
\end{figure}
It is hypothesised that the systematic is a function of $\Upsilon$, so that an additive correction factor to the raw genus curve can be defined
\begin{equation} 
 f(\nu;\Upsilon) = \sum_{n=1}^{n_\mathrm{max}} a_n(\Upsilon)\psi_n(\nu);
\end{equation}
for a given input power spectrum, the task is to empirically derive the correction coefficients $a_n(\Upsilon)$, which at best will require tabulation only to very small $n_\mathrm{max}$. This proves to be the case for power spectra that have been tried: the correction function is accurately described once  Hermite functions  up to fourth order are included, with no significant gains beyond this. Furthermore, above values of $\Upsilon\sim1$, the correction is in the first Hermite mode at the 95\% level, so that a formula can be given in terms of this quantity only. A figure showing the coefficient amplitude $a_1$ as a function of $\Upsilon$ is given in Figure~\ref{fig:sampling}. Extension to arbitrary $\Upsilon$ is achieved by interpolation or the use of the function:
\begin{equation} a_1(\Upsilon) = \beta_1\frac{1}{\Upsilon} + \beta_0;\quad \frac{a_1(\Upsilon)}{a_2(\Upsilon)} \simeq B_0\frac{1}{\Upsilon}\label{eq:upsilon};\end{equation}
in the example case presented here, the maximum-likelihood estimates of the parameters are $\beta_1 = -0.0548$ and $\beta_0 = 0.0166$, and $B_0 = 0.13$.
It is argued here that this formula is robust with respect to the choice of power spectrum, absolute sampling density and the resolution of the density field. Considering the form of the genus curve presented in Equation (2), it is well-known that the power spectrum of the Gaussian random field enters only in the constant of proportionality. For small departures from Gaussianity, such as those studied by Matsubara (2003), the genus curve is modified by low-order Hermite polynomials, as is the case here, with the coefficients of these polynomials depending on the power spectrum \emph{via} the skewness parameters. Crucially, because the Hermite mode coefficients $a_n$ are calculated only with respect to the shape of the curve and not to its amplitude, the leading-order influence of the power spectrum is removed.

The result is that the parameterised form of the correction presented in Equation (\ref{eq:upsilon}) is applicable to a wide variety of power spectra, though the parameters themselves will need to be tuned in each case. It is foreseeable that an analytic form of these parameters could be derived in the same fashion as the second-order perturbation theory analysis of non-Gaussianity arising from gravitational evolution, though this is beyond the scope of the current exposition; an attempt to demonstrate both the generality of this form and the derive a connection between the parameters and the underlying power spectrum would be a worthy future advance.

It is clear, however, that this correction is not useful at arbitrarily small $\Upsilon$. Indeed, while the prevailing rule of thumb that $\Upsilon\sim 1$ is sufficient to ensure a reasonable reconstruction seems to be borne out, there is nevertheless some correction required even then. While it is possible to go somewhat below this, the corrections quickly begin failing to converge; treatment when $\Upsilon < 0.75$ seems ill-advised---better to deem the field insufficiently reconstructed. Above this limit, however, the correction formula holds to better than 1\%. Further work will clarify the degree to which this correction holds for non-Gaussian (\eg, nonlinear) fields, exploring whether the small departures in Figure~\ref{fig:sampling} are a breakdown of this scheme.

\subsection{Selection function systematics}
An extension of the sampling systematic correction is its application to flux-limited galaxy samples in which the survey selection function is well understood. When the selection function is known, it can be used to upweight those locations in the density field that have been relatively under-observed (for whatever reason), prior to smoothing. Nevertheless, these regions will still have lower sampling rates, which will induce a sampling systematic that varies with location. To take the specific example of redshift, at high redshifts the average pair separation $\bar\ell$ will often be relatively higher even with the selection upweighting taken into account. Because the genus is a locally additive quantity (after the reconstruction has taken place), the formula (\ref{eq:upsilon}) can be used to correct the resulting measurement:
\begin{equation} f(\nu;\Upsilon) = \int_\mathcal{V} a_1\left[\Upsilon(\mathbf{x}) \right]d\mathbf{x}\end{equation} 
The improvement in signal-to-noise that results from using flux-limited samples depends on the specific selection properties of the survey, as the quality of the reconstruction of the density field is proportional to the number of galaxies used in its reconstruction; a volume-limited sample typically contains at most 10\% of the total sample, which propagates directly into the resulting genus signal (see, \eg, James et al. 2009). However, the type of sample that should be constructed will ultimately depend on the scientific application: a flux-limited sample will, in general, contain greater variation of the galaxy mass function with redshift, which will impede the interpretation of measurements. The purpose of the study of the sampling systematic is to ensure that, when the selection and galaxy sample properties are well understood, Gaussian smoothing reconstruction can be used to study scales close to the mean inter-object separation.

\subsection{Edge effect systematic}
The edge systematic has been treated in some depth by \citet{1993ApJS...86....1M}, who pioneered a windowing correction that modifies the reconstructed field before genus measurement, rather than correcting the measurement after it is made. The motivation for the existing method is that power will be pushed through these edges during the reconstruction, so that boundary regions should be expected to be systematically under-dense. This method works by creating a field of constant density value with the same geometry as the data, performing the same reconstruction procedure on it and ratio the smoothed data field by the smoothed constant field. This remains the best correction method for this effect, performing well for the current generation of surveys. These authors have noted that even when their correction method is applied, the measured genus curve should be scaled by a factor corresponding to the ratio of the total initial volume to the volume not pushed outside during the smoothing process, which is carried through to the computation of Hermite coefficients. As this affects all coefficients equally, their fractional contributions are unaltered.

\begin{figure}
\centering
\includegraphics[width=\linewidth]{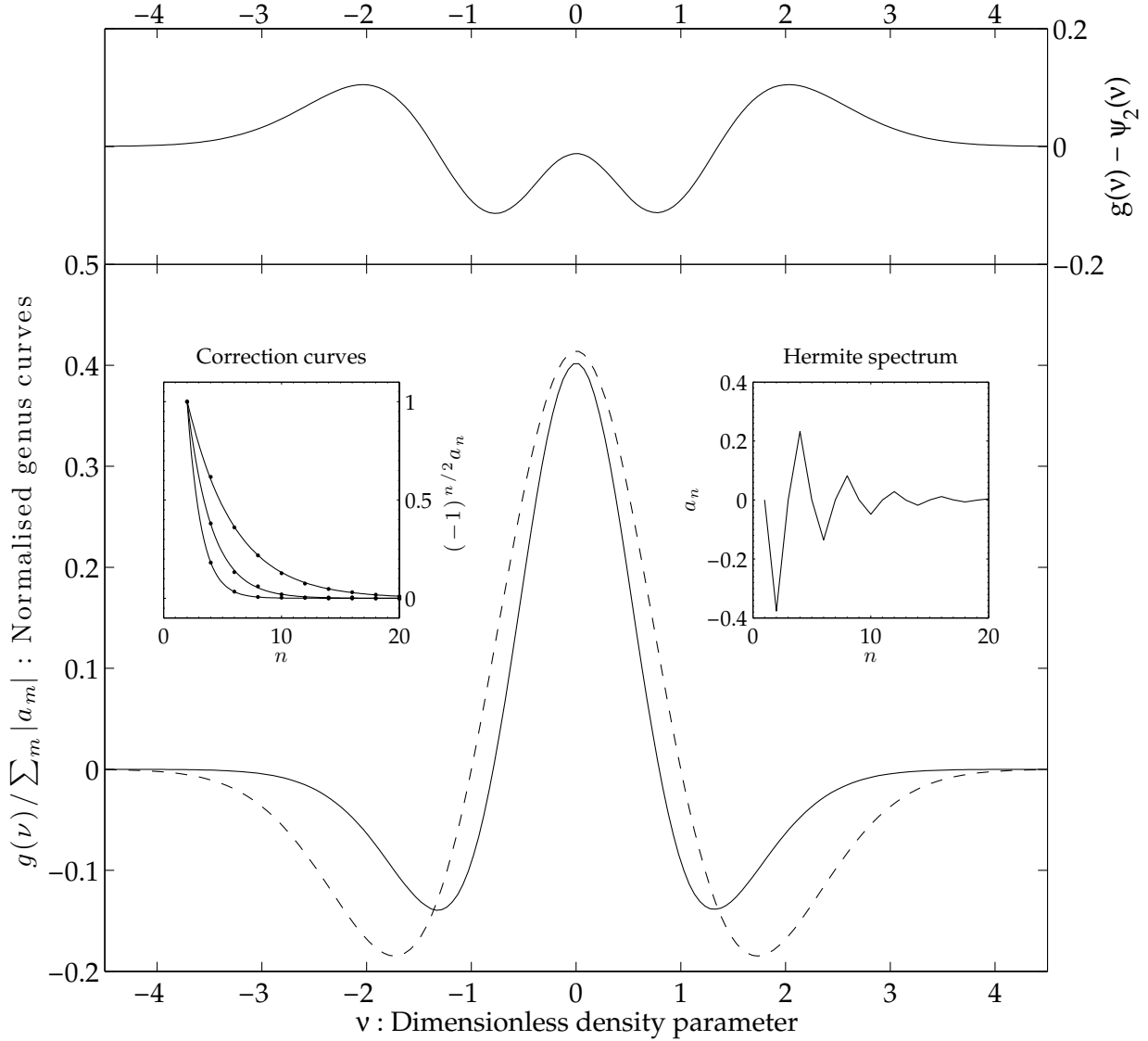}
\caption{Demonstration of the discretisation systematic effect, in which the genus curve of a pixellated Gaussian random density field (solid) is symmetrically distorted from its theoretical form (dashed), where each curve has been normalised so that the absolute sum of its Hermite coefficients is 1. The top panel shows the difference between the two curves; the right-hand inset shows the Hermite spectrum recovered from the decomposition of the measured genus curve, demonstrating that the systematic effect is not corrected by the treatment of only low-numbered modes; but, as in the left-hand inset, the formula for its correction obeys the relatively simple form of equation (\ref{eq:pixel_correct}), shown here with the data (points) and models (solid) for $\lambda\rightarrow0$ (no smoothing), $\lambda=0.5$ and $1.0$, tending to the limit of a delta function at $n=2$, corresponding to the Gaussian random field.}\label{fig:pixel}
\end{figure}
\subsection{Discretisation systematic}
Though the detailed specification of this systematic effect is unclear, it is known that a minimum smoothing length exceeding the scale of individual pixels must be applied to any pixellated density field. When this scale of smoothing is not met, the genus measurement is distorted symmetrically about median density in the manner shown in Figure~\ref{fig:pixel}. In work by~\citet{1986ApJ...309....1H}, the effect of finite (octahedral) pixel size was developed analytically, though without explicit invocation of a smoothing scale (\emph{cf.}~Equation (21) of that work). This computation shows that the expected expression (Equation (2) in the present work) is recovered in the limit of vanishing pixel size $r$ and that residual terms $\mathcal{O}(r^2)$ are well-described by quantities formulated from the higher spectral moments of the power spectrum. \citet{1989ApJ...345..618M} have developed this idea in the context of the two-dimensional genus statistic, but with application to cubical pixels of the kind most commonly employed in genus calculations. This impressive computation, to sixth order in the pixel size, shows clearly how polynomials in $\nu$, such as the Hermite polynomials, carried to progressively higher orders with coefficients expressed in terms of spectral moments, can account for the behaviour observed in the large pixel-size regime of genus computations.

We present here a treatment that complements the formalisms that have been developed previously, both by explicitly mixing consideration of the smoothing scale and by presenting a form of that correction that carries to all orders. For this systematic, the bias induced in the genus curve is not a function of a single Hermite mode, but of the full Hermite spectrum of the measurement. This complicated form (see right-hand inset in figure) suggests that treatment for this systematic can be achieved by the inclusion of corrections to Hermite coefficients at all orders. The proposed formula for this correction is
\begin{equation}
a_n = \left\{ \begin{array}{rl}
 0, & n=1,3,5,\ldots \\
 (-1)^{n/2}e^{-\gamma n} & n = 2,4,6,\ldots
\end{array} \right. ;\qquad
\gamma\approx \gamma_1n+\gamma_0, \label{eq:pixel_correct}
\end{equation}
where the maximum-likelihood parameter values are $\gamma_1=3.75$ and $\gamma_0=0.325$ for the field presented in this example. The left-hand inset in Figure~\ref{fig:pixel} demonstrates the variation of this systematic for several smoothing lengths ($\lambda = 0$, $0.5$ and $1.0$ pixels, tending toward the delta function $\delta(n-2)$ in the limit $\lambda\rightarrow\infty$. Even when $\lambda$ exceeds the pixel size, some effect remains. In practice, this systematic is not often an issue because the smoothing scale will naturally exceed the pixellation length by a sufficient factor in order to ensure that the sampling systematic is addressed, but further work can develop a treatment for cases where this is not achievable. 

A promising direction in which to extend this result is to consider expressing the coefficients of the correction in terms of the spectral moments of $P(k)$, as has been done in earlier work. While not attempted here, 
some guidance can be found by comparing the terms of the expansions presented in those previous results to the forms of Hermite polynomials, matching coefficients as polynomials of spectral moments.
 
\section{Application II: Astrophysical and cosmological phenomena}

\subsection{Nonlinear gravitational evolution}

\subsubsection{Perturbation theoretic treatment}
An influential formalism for the study of the topology at weakly nonlinear scales has been provided by \citet{Matsubara94} and developed energetically by \citet{2003ApJ...584....1M}. In this perturbative scheme, carried out in powers of the variance of the density field, the modes corresponding to $n=1, 3$ and $5$ appear with greater relative strength at scales below some scale, $\lambda\sim10$ Mpc$/h$ at the present epoch. This scheme is naturally re-expressed in terms of the current work, as a prediction for the spectrum of Hermite coefficients, yielding
\begin{equation}
g_\mathrm{NL}(\nu;\lambda) \propto e^{-\nu^2/4}\sum_{n=1}^5 a_n(\lambda)\psi_n(\nu) +\mathcal{O}(\sigma^2)
\end{equation}
with
\begin{equation}
\left[\begin{array}{c} a_1 \\ a_2 \\ a_3 \\ a_4 \\ a_5 \end{array}\right] =
\left[\begin{array}{c} S^{(2)}\sigma(\lambda)\\ \sqrt{2}\\  \sqrt{6}S^{(1)}\sigma(\lambda) \\ 0 \\ \sqrt{(10/3)}S^{(0)}\sigma(\lambda)\end{array}\right], \label{eq:ptng}
\end{equation}
where $S^{(0,1,2)}$ are the skewness parameters of~\citet{2003ApJ...584....1M} and $\sigma$ is the rms of the evolved density field at spatial scale $\lambda$. By expressing the non-Gaussian evolution in terms of the relative contribution of each mode, the result is independent of the power spectrum of the density field, which enters only as the overall normalisation of the genus curve and so scales all modes equally; all other coefficients are zero, to this order in the perturbation expansion.

These two approaches to non-Gaussian effects on cosmic structure topology lead naturally to the suggestion that this study be carried out in terms of the \emph{Hermite spectrum} of the genus curve. The ansatz for this programme is that the distinct processes inducing non-Gaussianity in the cosmological density field---survey selection effects, bias, redshift space distortions, nonlinear gravitational evolution and primordial physics---have differing spectral imprints that will be delineated. In the remainder of this communication, results applying this decomposition to the study of nonlinear evolution and selection effects are presented.

\subsubsection{Examination of Horizon Run simulation}
In this section, the formalism of Section 3 is applied to genus measurements in the Horizon Run of~\citet{Kim09}, a dark matter-only $N$-body simulation of some 70 billion particles across a volume of $L^3=(6.592 \textrm{ Gpc}/h)^3$, from which dark matter halos are drawn following the methodology of~\citet{Kim06}. The simulation volume studied here is a periodic cube at redshift zero, with halos of all masses down to resolved limit used, though, following convention, each object is assigned equal weight, rather than a weight proportional to its mass, during the reconstruction process.

The topology of structure in this simulation is examined across scales from 5 to 65 Mpc$/h$. At all scales, the spatial resolution of the density field and number of resolution elements in the measurement, are held fixed. To achieve this, subvolumes extending from the origin to $L_\mathrm{max}=100\lambda$ Mpc$/h$ are extracted from the full simulation and embedded within an array of size $N^3=400^3$. Periodicity is assumed during the density field reconstruction.

\begin{figure}
 \centering
 \includegraphics[width=\linewidth]{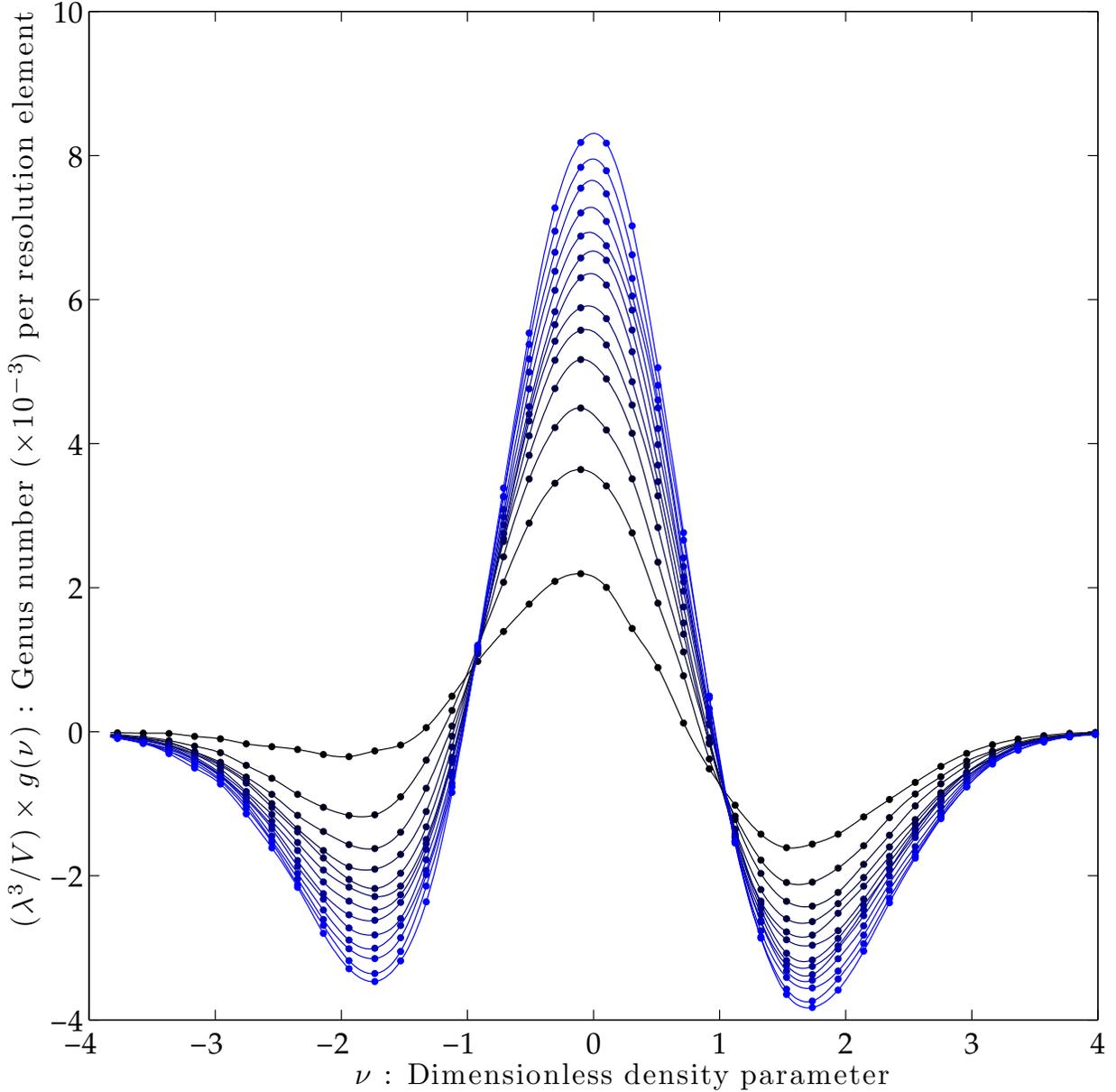}
 \caption{Genus curves for the Horizon Run subhalo population on scales of 5 (most black) to 65 (most blue) Mpc$/h$,  in steps of 5 Mpc$/h$, normalised to the number of smoothing kernels in the total survey volume. Asymmetry in the genus curve, most pronounced at the shortest scales where sampling density effects are relevant, nevertheless persists well into the linear regime (see Figure~\ref{fig:modes}). The mean inter-halo separation for the Horizon run is 15 Mpc$/h$; the corrected curve at 10 Mpc$/h$ may still be deemed useful, while the shortest scale presented in this plot is included as a demonstration of the behaviour of asymmetry in the strongly under-sampled limit.}\label{fig:genus}
\end{figure}

\begin{figure}
 \centering
 \includegraphics[width=\linewidth]{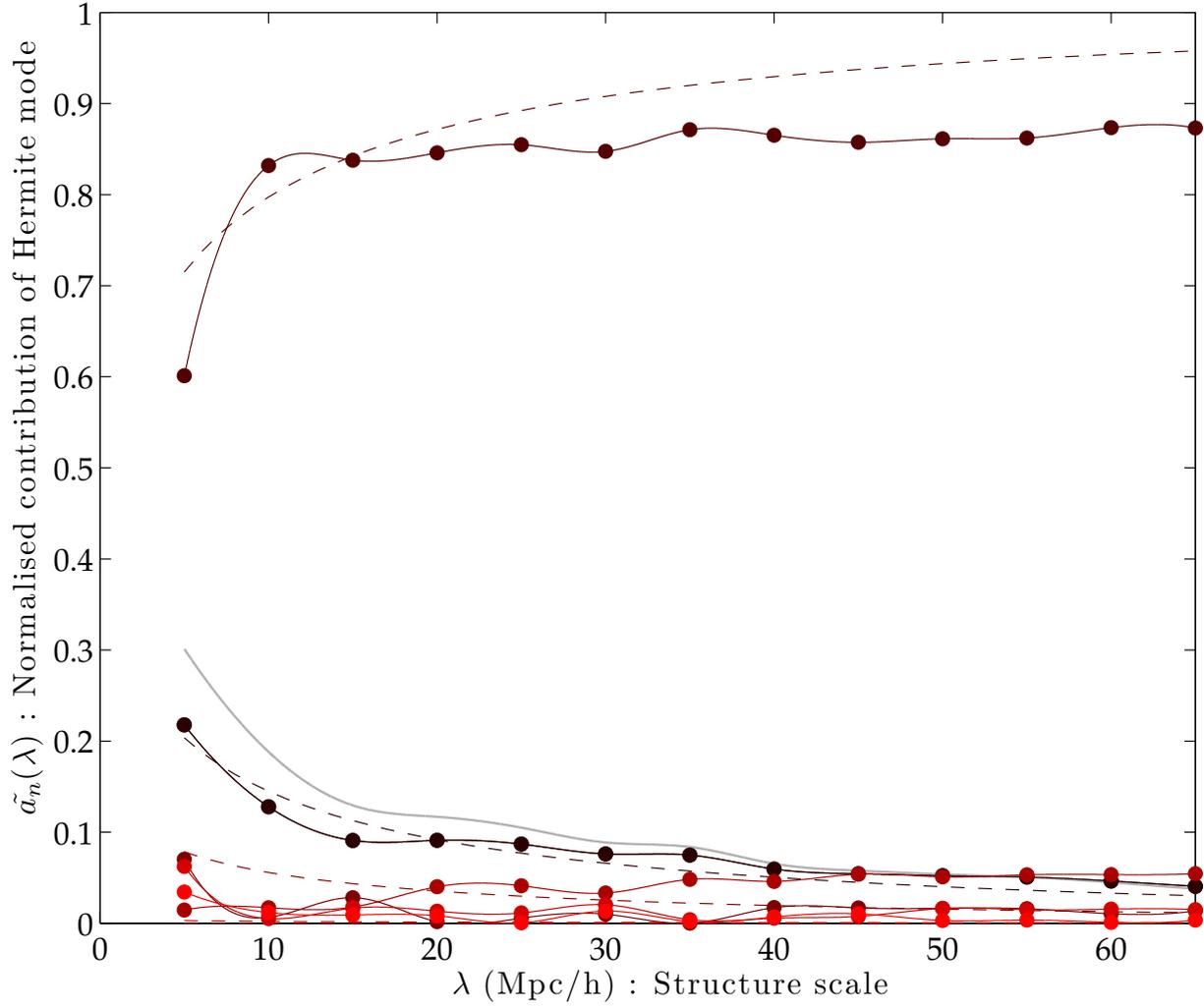}
 \caption{Hermite spectrum for the Horizon Run genus curves in Figure~\ref{fig:genus} as a function of scale: fractional contribution of modes $a_n$ from the nonlinear to linear regime (points and solid lines, with the mode $n=1$ in black), compared to the predictions from second-order perturbation theory (dashed). The sampling correction of equation (\ref{eq:upsilon})  has been applied to the $n=1$ mode; the uncorrected mode is shown in grey. }\label{fig:modes}
\end{figure}

Figure~\ref{fig:genus} shows the genus curves measured from the Horizon Run subhalo catalogue on scales ranging from 5 to 65 Mpc$/h$. It is surprising to see an asymmetry toward isolated high-density structures even at the largest scales of reconstruction, which is well into the linear regime; this result confirms the analysis carried out by~\citet{Kim09}, which is in contrast to earlier studies. To explore this measurement further, the Hermite decomposition of the genus curves is presented in Figure~\ref{fig:modes}, in which the contribution from each mode is normalised to give its fractional contribution to the total: $\tilde a_n = |a_n| / \sum_m |a_m|$.  The decomposition expresses concisely the degree of nonlinear evolution in the density field as reconstructed from these objects. It is to be expected that part of the non-Gaussian signal in the decomposition arises from the use of biased tracers of the density field, though this is unlikely to manifest itself in a scale-dependent fashion at the scales studied here in the absence of any primordial non-Gaussianity, so that the relative evolution of the odd-numbered Hermite modes (Figure~\ref{fig:modes}) is taken to be a description of the modification to topology induced by nonlinear gravitational evolution. At the shortest scales, the sampling density falls below that required for direct reconstruction of the field, so that the systematic correction derived in the previous section is applied to the measurement. Even with this correction applied, it is clear that the mode corresponding to a pure Gaussian random field is under-represented relative to the expectation from second-order perturbation theory, particularly evident at the largest scales.

\subsection{Redshift-space distortions}
The peculiar motion of galaxies affects the genus measurement only through its impact on the reconstructed density field. No longer a systematic to be overcome, this coupling between the density and velocity fields yields information about the processes of gravity and structure formation. The linear treatment of this coupling, the Kaiser limit, has been shown to generate no distortion to the shape of the genus measurement for a Gaussian random field; like the survey boundary correction, this affects the amplitude of the genus curve, so that the fractional contribution from Hermite modes is unchanged. \citet{Matsubara96} has speculated that the short scale distortion effects, which act orthogonally to those in the Kaiser limit, will also alter just the amplitude of the curve, a claim that has been examined in some simulations~\citep{2005ApJ...633....1P}. Given the interest directed toward redshift-space phenomena in the field at present, it seems plausible that in the intermediate limit between these two regimes could host effects that do alter the shape of the genus curve, though it is not clear that such scales can be examined in the current generation of redshift surveys.

\subsection{Primordial non-Gaussianity}
Broad classes of inflationary models generate significant non-Gaussianity in the early-time cosmological density field. The customary formalism for this effect is through the (Bardeen) curvature perturbation in the matter era:
\[ \Phi = \phi + f_\mathrm{NL}(\phi^2 - \langle\phi\rangle^2), \]
where $\phi$ is a purely Gaussian random variable and $f_\mathrm{NL}$ is taken to characterise the magnitude of non-Gaussianity in the field~\citep{KS01}.  The bispectrum of the primordial curvature perturbation, identically zero in a Gaussian random field, will peak in regions of Fourier space for which the closed triplet of wavenumbers $\lbrace \mathbf{k}_1, \mathbf{k}_2, \mathbf{k}_3\rbrace$ exhibits one of a handful of ratios, giving rise to a classification of primordial non-Gaussianity in terms of triangle configurations. Of concern here are the `squeezed' (\emph{cf}.~`local') ($k_1 \ll k_2 \sim k_3$), equilateral ($k_1\sim k_2 \sim k_3$) and enfolded ($k_1 \sim k_2 \sim k_3/2$) configurations, corresponding respectively to inflations scenarios involving multiple fields, horizon crossing and approximate shift symmetry in very general single-field models~\citep{Wagner10}.

A study of the influence of primordial non-Gaussianity on the topology of large-scale structure has been undertaken on the scales above 100 Mpc$/h$ by~\citet{HKM}, who present formulae for the genus statistic (among others) from perturbation theory. In comparing these analytic results to $N$-body simulations, they note that the impact of primordial non-Gaussianity should be slight in proportion to the degree of gravitationally induced non-Gaussianity in the field. Nevertheless, the discovery by~\cite{Dalal08} of a scale-dependent galaxy bias induced by local-type primordial non-Gaussianity provides sufficient motivation to attempt an exploration of this effect through the formalism presented in this work.

Here, the impact of primordial non-Gaussianity on the topology of the late-time density field has been evaluated empirically with the simulation data set of \citet{Wagner10}. These data are the output of a sequence of $N$-body simulations with generic non-Gaussian conditions, where the departure from Gaussianity is characterised completely by the bispectrum of the  initial conditions.  Power spectra, halo mass functions and bias properties of these simulation data have been presented by~\citet{Wagner10} and \citet{Wagner11}. In this suite, halo catalogues derived from an evolved density with initial non-Gaussianity $f_\mathrm{NL} = \lbrace0,\pm100,\pm250,\pm500\rbrace$ have been produced, in each of the local, equilateral and enfolded configurations. Snapshots have been provided at $z\sim\lbrace0,0.33,0.67,1,1.5,2\rbrace$; in this work, we consider halo catalogues from the redshift one density field with a view to optimizing the number density of halos and (reciprocally) the degree of nonlinear evolution in the underlying density field. Concurrent work by Wales, James \& Lewis (2012) incorporates an examination of the redshift dependence of these measurements.

The halo catalogue output of these simulations is transformed into a counts-in-cells density field in the same manner as the data in the previous section. To evaluate the departures from non-Gaussianity present in the genus measurement of these data, the magnitude of the base $(n=2)$ mode of the decomposition, relative to that of the Gaussian case, is studied across the range of scales from 15 to 65 Mpc$/h$, in steps of 5 Mpc. The scale at which $\Upsilon = 1$ is around 12 Mpc, so all scales examined are free from the impact of the sampling systematic. Genus curves are measured for each ($f_\mathrm{NL}$, configuration) pairing, in each of the eight available simulations, before the Hermite decomposition is performed on each measurement. The variance between the measurements of the Hermite coefficients for the eight simulations is taken to be an estimator of the variance of the coefficient statistic.

\begin{figure}
\centering
\includegraphics[width=\linewidth]{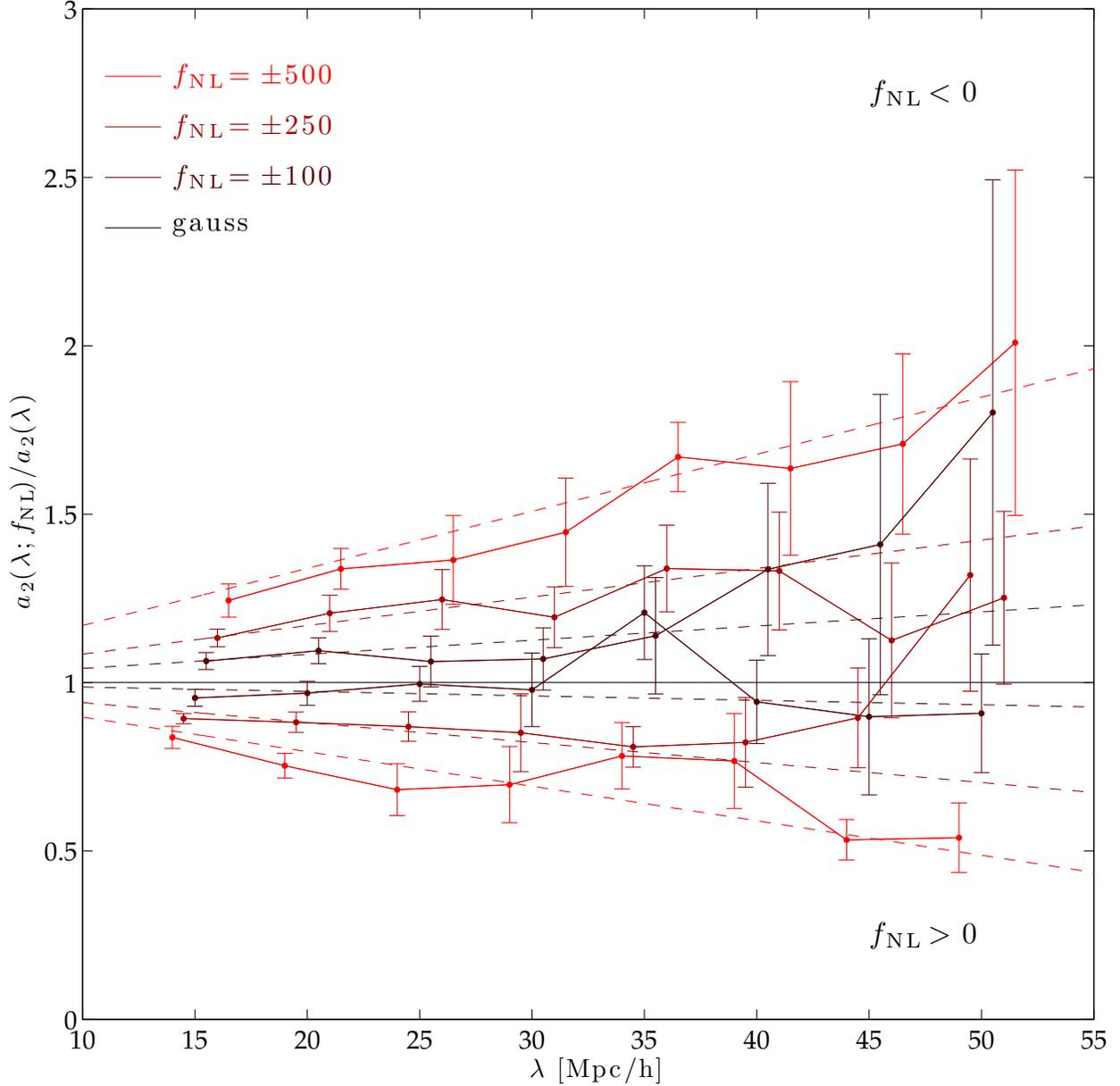}
\caption{The scale dependence of the $a_2$ coefficient of the genus curve for varying $f_\mathrm{NL}$ in the local configuration relative to the Gaussian ($f_\mathrm{NL}$) case, averaged over eight simulation realisations (solid), with 1$\sigma$ error bars shown. The points are displayed in a staggered manner at each scale for clarity. The maximum likelihood fit of a one-parameter model of linear scale dependence (dashed) indicate a relationship between the rate of scale dependence and $f_\mathrm{NL}$.}\label{fig:NGsims_fNL}
\end{figure}
\begin{figure}
\centering
\includegraphics[width=\linewidth]{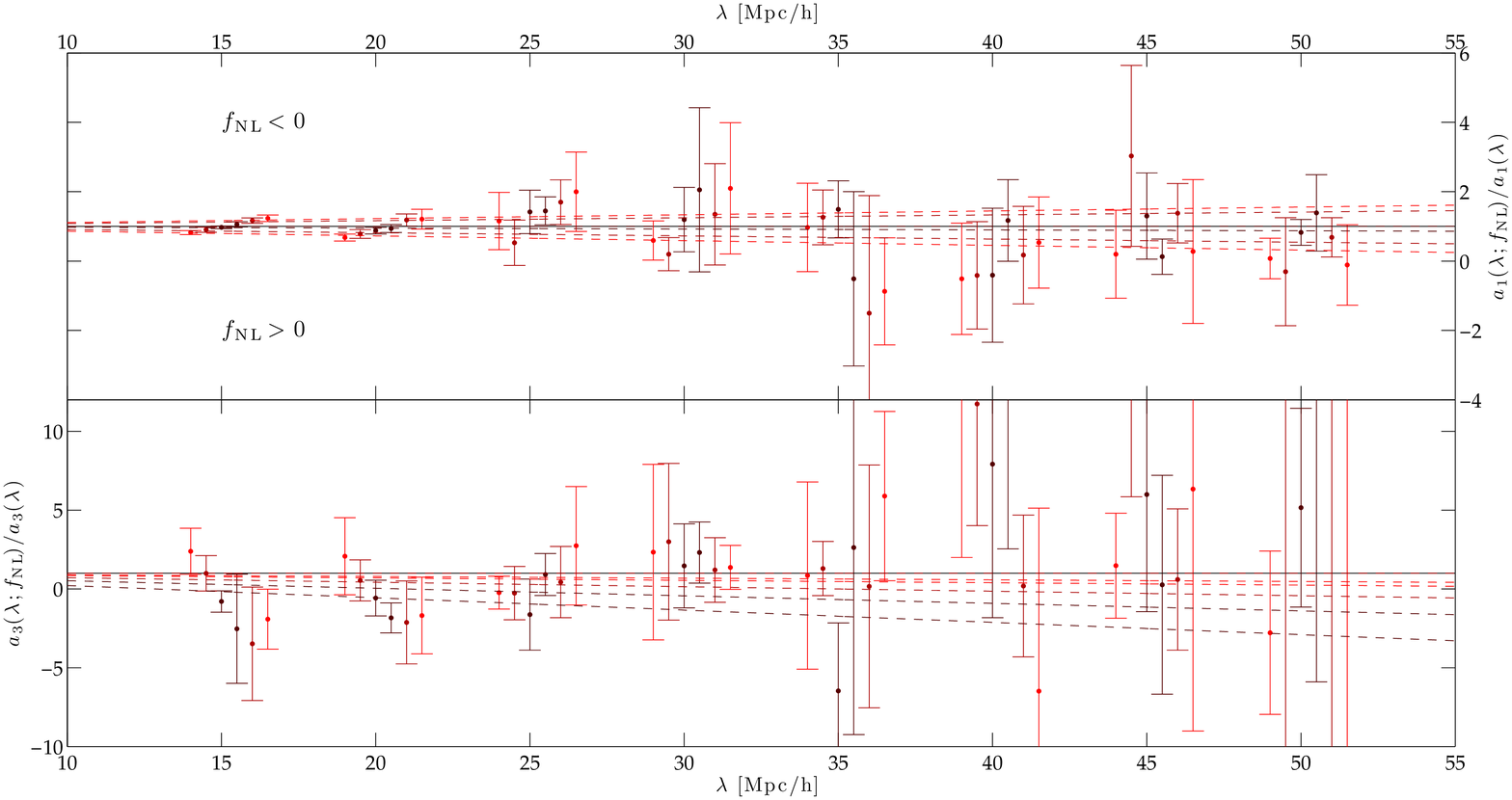}
\caption{Empirical demonstration of the diminshed impact of $f_\mathrm{NL}$ primordial non-Gaussianity on the weaker modes ($n=1$ and $n=3$) of the redshift zero genus curve, a result that appears to hold across the scales available in the simulation set used in this work, with the same labelling scheme as Figure~\ref{fig:NGsims_fNL}. The diagonal elements of the covariance matrix, shown as $1\sigma$ error bars, have been offset for clarity at each scale.}\label{fig:modes_fNL}
\end{figure}

Figure~\ref{fig:NGsims_fNL} shows the relationship between $f_\mathrm{NL}$ and the relative-to-Gaussian strength of the $n=2$ Hermite mode. This is an interesting result that has several possible interpretations. Perhaps the most satisfactory is that, as the amplitude of the genus curve is a functional of the (biased) power spectrum, even though the underlying dark matter power spectrum is fixed for a given redshift, configuration and realisation, the scale dependence of the quantity $a_2(f_\mathrm{NL})/a_2(\textrm{gauss})$ is a measure of scale-dependent bias. A prediction of this hypothesis is that all other such ratios $a_n(f_\mathrm{NL})/a_n(\textrm{gauss})$ should show the same effect, however these data do not allow such a test to be carried out at sufficient significance; e.g., Figure~\ref{fig:modes_fNL}) shows the same measurement as Figure~\ref{fig:NGsims_fNL} with the lower significance modes of the genus curve. 
\begin{figure}
\centering
\includegraphics[width=\linewidth]{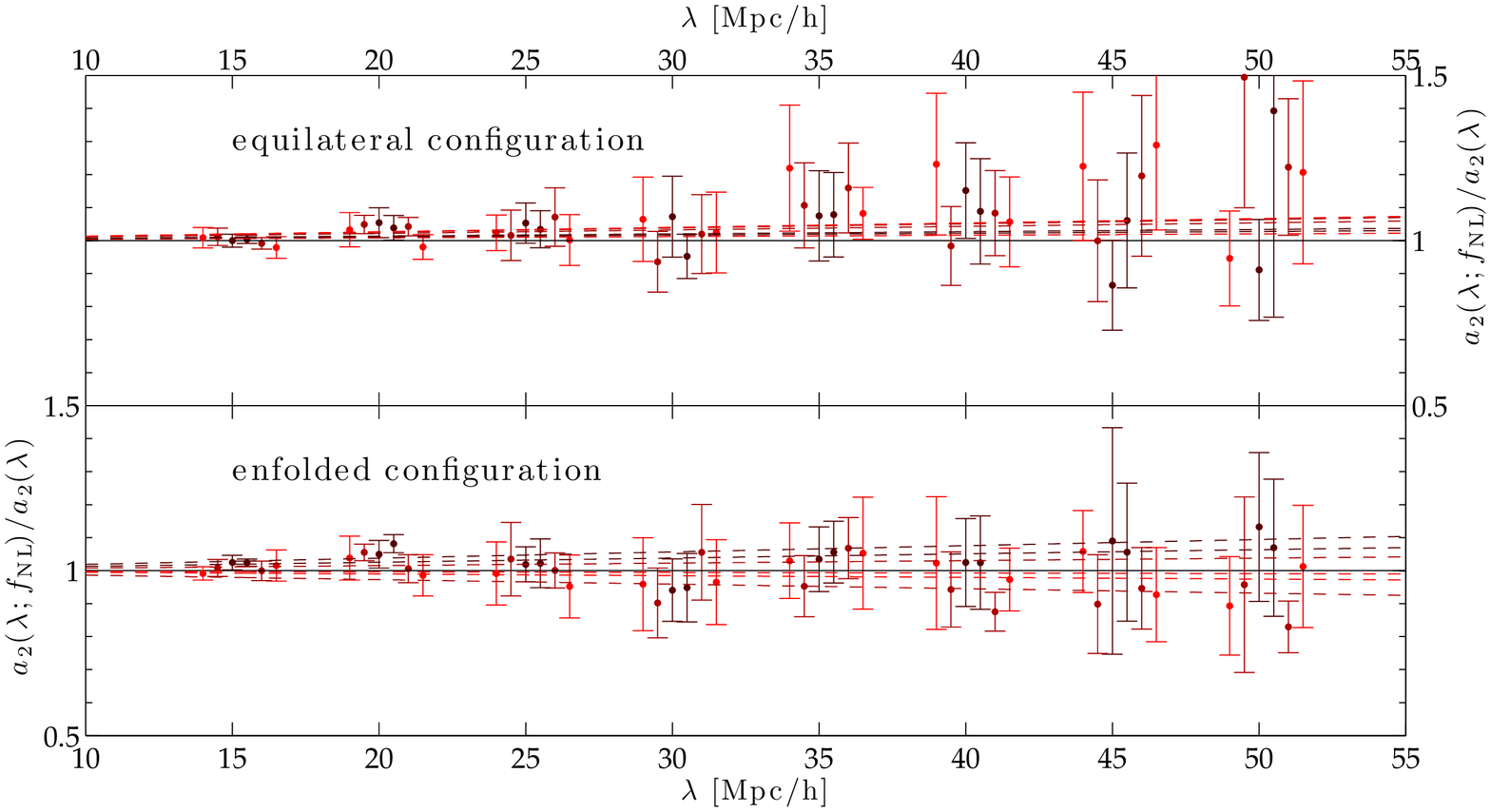}
\caption{Same measurement as Figures~\ref{fig:NGsims_fNL} and \ref{fig:modes_fNL}, for alternative configurations of the initial bispectrum. In contrast to the local configuration, no scale-dependent bias is apparent in these results.}\label{fig:modes_NGconf}
\end{figure}
In this hypothesis, it is significant that the local configuration shows this scale dependence. Figure~\ref{fig:modes_NGconf} shows how the other configurations of the bispectrum do not result in the same behaviour. In some models it is predicted that the equilateral configuration should show a $1/k$ scale-dependent bias, which would be expected to manifest here. Even though this is weaker than the $1/k^2$ dependence of the scale-dependent bias in the local configuration, it is somewhat surprising that only the local configuration displays this behaviour---however, this seems to be compatible with the results of~\citet{Wagner10}.

\begin{figure}
\centering
\includegraphics[width=\linewidth]{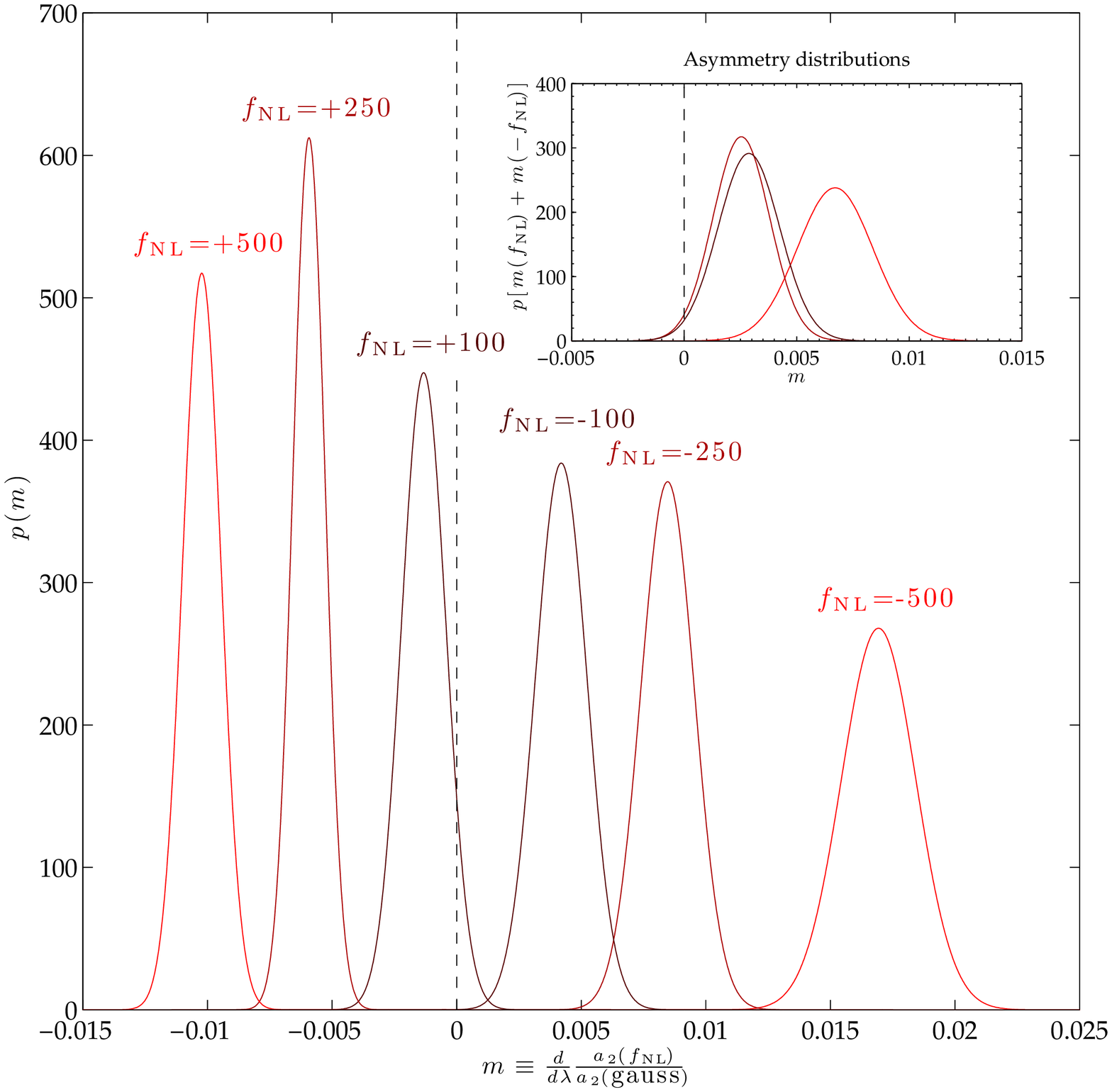}
\caption{Assuming a one-parameter linear relationship between scale and $a_2(f_\mathrm{NL})/a_2(\textrm{gauss})$, the likelihood distribution of the gradient $m$ is shown for each of the $f_\mathrm{NL}$ values studied in this work (solid curves, main plot). Assuming that $m$ is a normally distributed variable for each $f_\mathrm{NL}$, it is apparent that the genus statistic is \emph{not} symmetrically sensitive to $f_\mathrm{NL}$ at the 2$\sigma$ level for $|f_\mathrm{NL}|=\lbrace 100,250\rbrace$, and at the 4$\sigma$ level for $|f_\mathrm{NL}|=500$ (inset).}\label{fig:NGsims_fNL_likelihoods}
\end{figure}

An alternative explanation that is tractable by the means presented in this work is that, if the bispectrum manifests in the genus curve in a manner that is not simply related to the amplitude, then the results in Figure~\ref{fig:NGsims_fNL_likelihoods} constrain the mechanism of this action. It might be hoped that, given the exact prediction for the genus curve of a Gaussian random field, many distinct types of non-Gaussianity could be examined, even phenomenologically. This is a topic that requires a wider set of simulation data to examine and that will be explored in the future.

To return briefly to the hypothesis of scale-dependent bias manifesting in the amplitude of the genus curve: recent work by several authors has given rise to methods of constraining dark energy and modified gravity by means of the genus curve amplitude~\citep{2010arXiv1010.3035W,2011MNRAS.412.1401Z}, based on the development of a standard ruler from large-scale structure by~\citet{2010ApJ...715L.185P}. This class of tests probes the cosmological parameters via the linear regime density field as well as the comoving volume element, both of which determine the genus amplitude. Cosmological density fields that possess some primordial non-Gaussianity will also preserve their initial (non-Gaussian) topology into the linear regime. The impact of primordial non-Gaussianity, as measured and interpreted in the present work, can be used as part of the standard ruler technique, by exposing an additional scale-dependent term (that is, beyond the usual scale dependence of the genus curve amplitude) that does not manifest within the comoving volume element. This improves the possibility of distinguishing these phenomena using the genus curve.

\section{Discussion}
The challenge posed by this result is to delineate distinct physical processes in terms of their impact on particular Hermite modes (or, more generally, sequences of such modes). This article has introduced a Hermite decomposition of the genus curve for cosmic structure, Eq.~(\ref{eq:decomposition}) that serves as a pivot between genus measurements, physical modelling such as that from second-order perturbation theory and the systematic effects that arise in the measurement of the genus statistic in galaxy redshift surveys. In contrast to existing genus-related statistics, because the set of coefficients $a_n$ is based on an orthogonal sequence of polynomials, the error analysis of this measurement is greatly simplified; however, the rather natural interpretation of $\Delta\nu$, $A_c$ and $A_v$ as corresponding to structures at different densities has been lost in this scheme.

This technique has been used in the analysis of the WiggleZ galaxy redshift survey, the results of which are published in a forthcoming work~\citep{James12}. An important consideration in this programme is the density resolution required for a given level of accuracy in the decomposition: measuring higher Hermite mode coefficients requires that the genus curve be well sampled in $\nu$. While individual genus measurements are relatively inexpensive to compute, large gains are realised when GPU computation is used, a topic explored in contemporary work by~\citet{Wales12}. Analysis of this sort is well-matched to immediate use in the current generation of surveys, extending their capacity for description of non-Gaussian cosmic structure. 

The aspiration of this research programme is to provide an analytical account of the relationship between the physics behind non-Gaussian fields and the topological properties of the cosmic structures that arise as a results. Because the theoretical understanding of the relationship between some physical phenomena, particularly galaxy bias and nonlinear regime redshift-space distortions, is not established at this point, the methods presented in this paper cannot yet be exploited in full. Table~\ref{table:dists} summaries this situation by describing these phenomena through the distribution of Hermite modes that result from the decomposition. It would be reasonable to expect that, in time, the impact of redshift-space distortions, galaxy bias and perhaps more general models for primordial non-Gaussianity, can display more complex behaviour under the decomposition, which would improve the prospect of understanding them through the topology of large-scale structure.
\begin{table}
\centering
\begin{tabular}{ccc}
\hline\hline
Effect & Distribution & Comments\\ \hline
\textit{Systematic selection effects} & &\\
Sampling systematic & Delta ($n=1$) & \\
Finite-pixel size effect & Geometric & \\
Edge effects & Uniform &\\
\hline
\textit{Physical phenomena} & & \\
Nonlinear gravitational evolution & - & weakly nonlinear regime\\
Redshift-space distortions & Uniform & linear regime\\
Primordial non-Gaussianity & Uniform & assuming scale-dependent bias\\
$f(R)$ modified gravity & Uniform & \\
Dark energy density & Uniform & \\
\hline\hline
\end{tabular}
\caption{Distributional summary of results: interpreting the normalised spectrum of Hermite coefficients $\tilde a_n$ as a probability distribution, each of the physical and systematic selection effects discussed in this work, correspond to distributions that act multiplicatively on the Hermite spectrum. The current theoretical description of the effects of linear regime redshift-space distortion is that it alters the amplitude of the genus curve, so that the distribution of Hermite coefficients is uniform; other physical phenomena, with the exception of nonlinear gravitational evolution, act on the genus curve through the power spectrum and/or comoving volume element, though the short scale impact of these effects has not yet been studied.}\label{table:dists}
\end{table}

\clearpage



\acknowledgments

I should like to thank Alan Heavens and John Peacock for comments on this work and Fergus Simpson for making explicit the definition of $f_G$ presented in Eq (1). An anonymous referee suggested valuable improvements to the manuscript and encouraged its development. I acknowledge the efforts of the Horizon Run and Barcelona non-Gaussian structure simulators and laud their decision to make these data public.






\appendix

\section{Consolidation with prevailing statistics}
This Appendix presents a partial reconciliation between the formalism presented here and prevailing derived statistics of genus measurements. Beginning with~\citet{1992ApJ...392L..51P}, the degree to which a measured genus curve departs from the standard form of Equation (\ref{eq:grf}) has been quantified by three numbers, termed ``genus-related statistics'', that are special cases of the function
\begin{equation}
 G(a,b,m) \equiv \frac{\int_a^b \nu^m g(\nu) d\nu}{\int_a^b g_\mathrm{GRF}(\nu) d\nu};
\end{equation}
in particular
\begin{equation}
\Delta\nu = G(-1,1,1);\;A_c = G(1.2,2.2,0);\;A_v = G(-2.2,-1.2,0)
\end{equation}
are respectively the genus shift, cluster overabundance and void overabundance parameters. It is unfortunate that a little effort is necessary to recast this standard set of non-Gaussian descriptors in terms of the present work. A slight modification permits exploitation of a Hermite function recurrence relation
\begin{equation}
\int e^{-\nu^2/4} g(\nu) d\nu = -e^{-\nu^2/4}\sum_{n=1}^\infty \frac{a_n}{\sqrt{n}}\psi_{n-1}(\nu),
\end{equation} suggesting
\begin{equation}
G'(a,b,m) \equiv \frac{\int_a^b e^{-\nu^2/4}\nu^m g(\nu) d\nu}{\int_a^b e^{-\nu^2/4}g_\mathrm{GRF}(\nu) d\nu}, 
\end{equation} so that
\begin{equation}
A_c' = \left[ {\sum_{n=1}^\infty\frac{a_n}{\sqrt{n}}\psi_{n-1}(\nu)}\middle/{\frac{\psi_1(\nu)}{\sqrt{2}}}\right|_{\nu=1.2}^{2.2} 
\end{equation} and similarly for the void overabundance. An expression of $\Delta\nu'$ might also be derived, though it is regrettably difficult to reconcile the two sets of descriptors. Nevertheless, the study of deviations from Gaussianity across density ranges corresponding to clusters, filaments or voids is recast as a straightforward evaluation of Hermite functions at the density threshold values that are to be examined; it is stressed that the modification required to facilitate this has no bearing on the underlying physics.

\end{document}